\documentclass[preprint,amsmath,amssymb,aps,showpacs,superscriptaddress,nofootinbib]{revtex4-1}
\usepackage{epsfig,amssymb,amsmath,latexsym}
\usepackage{pifont}
\usepackage{hyperref}

\usepackage{color}
\definecolor{nicered}{rgb}{0.7,0.1,0.1}
\definecolor{nicegreen}{rgb}{0.1,0.5,0.1}

\newcommand{\be}{\begin{equation}}
\newcommand{\ee}{\end{equation}}
\newcommand{\bea}{\begin{eqnarray}}
\newcommand{\eea}{\end{eqnarray}}
\newcommand{\beal}{\begin{aligned}}
\newcommand{\eeal}{\end{aligned}}

\newcommand\ba{\begin{array}}
\newcommand\ea{\end{array}}

\renewcommand\sin{\text{sin}}
\renewcommand\cos{\text{cos}}

\newcommand\SEC[1]{\medskip\noindent{\sl\bfseries #1}}

\begin{document}
\preprint{DCPT-19/09}
\title{Accelerating Black Hole Chemistry}

\author{Ruth Gregory}
\affiliation{Centre for Particle Theory, Department of Mathematical Sciences
and Department of Physics, Durham University, South Road,
Durham, DH1 3LE, UK}
\affiliation{Perimeter Institute, 31 Caroline St., Waterloo, Ontario,
N2L 2Y5, Canada}
\author{Andrew Scoins}
\affiliation{Centre for Particle Theory, Department of Mathematical Sciences
and Department of Physics, Durham University, South Road,
Durham, DH1 3LE, UK}

\date{\small \today}

\begin{abstract}
We introduce a new set of chemical variables for the accelerating black
hole. We show how these expressions suggest that conical defects
emerging from a black hole can be considered as true hair -- a new charge
that the black hole can carry -- and discuss the impact of conical deficits
on black hole thermodynamics from this `chemical' perspective. We
conclude by proving a new {\it Reverse Isoperimetric Inequality} for 
black holes with conical defects.
\end{abstract}


\maketitle

Black hole thermodynamics has seen a rich renaissance over the past
decade, as the understanding of the role of thermodynamic pressure
has been fleshed out and explored \cite{Teitelboim:1985dp,Kastor:2009wy,
Dolan:2010ha,Cvetic:2010jb,Dolan:2011xt,Kubiznak:2012wp,
Dolan:2013ft,Kubiznak:2016qmn}. 
A key conceptual development was understanding that the mass term 
in the First Law of thermodynamics, related to the
mass parameter $m$ in the Newtonian potential of the black hole, was
not in fact the internal energy of the black hole, but rather its enthalpy
\cite{Kastor:2009wy},
i.e.\ the natural first law for a black hole with a cosmological constant
includes not only the charges of electromagnetism and rotation, but
also the impact of the non-zero energy coming from the cosmological
constant in the volume inside the black hole:
\be
dM = TdS + VdP + \Omega dJ + \Phi dQ\,.
\label{firstlaw}
\ee

Once one includes the possibility of a varying pressure, black hole
thermodynamics more naturally resembles conventional thermodynamics,
not only in its differential sense, but also in its integrable sense: the 
ideal gas relations $dU = TdS - PdV$, $U = c_V PV$ have their counterpart
in the differential first law \eqref{firstlaw} and an integral Christodoulou-Ruffini 
\cite{Christodoulou:1972kt,Caldarelli:1999xj} relation, that for four-dimensional
Kerr-Newman-AdS black holes reads
\be
M^2 \! = \frac{S}{4 \pi } \left [ 1 + \frac{\pi Q^2}{S} +\frac{8PS}{3}\right]^2
+ \frac{4\pi^2 J^2}{S}\left[1+\frac{8PS}{3}\right]\,,
\label{ChrisRuf}
\ee
and can be massaged into an ideal-gas like relation at large 
volume/entropy. Although the formulae for the
enthalpy, charges and potentials are naturally derived from the
black hole geometry, and written in terms of the metric parameters
and horizon radius, expressing the thermodynamic 
potentials and enthalpy purely in terms of extensive quantities
(such as explored in \cite{Dolan:2011xt,Dolan:2013ft})
allows a natural identification with classic thermodynamics, and
elucidates the \emph{chemical} nature of the phase space of
black holes. 

There is another reason to specify the closed form of the black hole
charges and enthalpy: While it is possible for material systems to have 
many charges and chemical potentials, the black hole on the other hand is 
believed to carry only mass, charge and angular momentum,
a situation summed up by the \emph{no-hair theorems} 
\cite{Ruffini:1971bza,Chrusciel:2012jk}, and while
these are now understood in a broader context to be somewhat limited,
the basic picture from the perspective of classic black hole
thermodynamics was that thermodynamic charges were still
narrowly restricted, $M=M(S,P,J,Q)$. Recently however, a new type of
`charge' for a black hole has been explored and added to this stable:
a conical deficit, $\mu$ \cite{Appels:2016uha,Appels:2017xoe,
Gregory:2017ogk,Anabalon:2018ydc,Anabalon:2018qfv}, 
often interpreted as a cosmic string, that can
either run symmetrically along the axis of the black hole
\cite{Aryal:1986sz,Achucarro:1995nu,Gregory:2013xca,Gregory:2014uca},
or have different values along the North and South axes, leading to
an \emph{accelerating} black hole, encoded by the C-metric
\cite{Kinnersley:1970zw,Plebanski:1976gy} (and including a negative 
cosmological constant $\Lambda = -3/\ell^2$):
\be
\beal
ds^2 = \frac{f(r)}{\Sigma H^2}\Big[
\frac{dt}{\alpha}-a\,\sin^2\theta \frac{ d\varphi}{K} \Big]^2 
\!- \!\frac{\Sigma dr^2}{f(r)H^2} \; - \frac{\Sigma r^2}{g(\theta)H^2}d\theta^2
\!-\! \frac{g\sin^2\theta}{\Sigma r^2 H^2} \Big[\frac{adt}{\alpha}-(r^2+a^2)
\frac{d\varphi}{K}\Big]^2
\eeal
\label{eq:metric}
\ee
where the usual Schwarzschild potential $1-2m/r$ is augmented not only by 
the charge and angular momentum terms, but also by an ``acceleration''
parameter, $A$:
\be
\beal
f(r) &=(1-A^2r^2)\left[1-\frac{2m}{r}+\frac{a^2+e^2}{r^2}\right]
+\frac{r^2+a^2}{\ell^2}\\
g(\theta) &=1+2mA\cos\theta+ (\Xi-1)\cos^2\theta\,,\\
\Sigma &=1+\frac{a^2}{r^2}\cos^2\theta\,, \qquad
H=1+Ar\cos\theta \,,\\
\Xi &= 1 + e^2 A^2 - \frac{a^2}{\ell^2} (1-A^2 \ell^2)
\eeal
\ee

Note that the black hole is assumed to spin on its axis, the 
acceleration term modifying the angular parts of the metric, 
distorting the sphere to a teardrop, with a conical deficit (at least) at 
one of the poles. This deficit is revealed by taking the limit of the metric 
as we approach each pole, and is encoded by the {\it tension} 
\be\label{eq:deficits}
\mu_\pm =\frac{1}{4}\bigg[1-\frac{\Xi \pm 2mA}{K}\bigg]\,.
\ee
(with `$+$' corresponding to the North Pole, and `$-$' the South)
interpreted as a cosmic string emerging from the black hole,
causing it to accelerate.

Usually, an object under uniform acceleration will have an
acceleration horizon, as ultimately an accelerating object
will asymptote the speed of light, however, in AdS spacetime,
the negative curvature of the space means that an object at fixed
finite displacement from the centre of the spacetime is actually
undergoing uniform acceleration. This means that a black hole
suspended from the boundary by a cosmic string is indeed
accelerating, even though the sole horizon is that of the black hole.
This r\'egime of accelerating black hole solutions that are truly
static with respect to an observer at the boundary are called
\emph{slowly accelerating} black holes \cite{Podolsky:2002nk}.
Roughly speaking, the criterion for slow acceleration is that the
scale set by acceleration, $A^{-1}$ is much larger than the 
AdS radius, $A\ell<1$,
although the true limit is dependent on the mass, charge and
angular momentum of the black hole. As the acceleration increases,
the position of the suspended black hole moves closer to the boundary
until at $A\ell\sim1$ there is a shift in the global structure of the spacetime
and for $A\ell>1$, the black hole now accelerates in from, and out to, 
the AdS boundary, see \cite{Podolsky:2003gm} for a discussion of 
the causal structure of the C-metric.

In a series of papers, \cite{Appels:2016uha,Appels:2017xoe,Gregory:2017ogk,
Anabalon:2018ydc,Anabalon:2018qfv} (see also \cite{Astorino:2016xiy,
Astorino:2016ybm}) the thermodynamics of conical deficits and accelerating
black holes was explored and refined. The key insight was to use tools from
holographic renormalization to properly calculate the various
charges of the slowly accelerating black hole spacetime 
\cite{Anabalon:2018ydc,Anabalon:2018qfv}. 
The nett result is a set of thermodynamic
variables for the black hole, expressed in terms of the black hole 
metric parameters and the horizon radius $r_+$, that include the conical
deficit as a charge, and introduce the conjugate chemical potential, a
{\it thermodynamic length}. The First Law has full cohomogeneity:
\be\label{flaw}
d M=T d S+\Phi d Q+\Omega d J+\lambda_+d \mu_+
+\lambda_-d \mu_- +Vd P\,,
\ee
with all the physical parameters of the geometry corresponding to 
a thermodynamic charge.
One of the key difficulties in determining the
correct enthalpy was in properly identifying the timelike Killing vector for
determining the mass of the black hole. The slowly accelerating black hole,
being at a fixed point from the boundary, has the same time coordinate
(up to a factor) as the asymptotic AdS spacetime, and the mass can be 
found via a holographic renormalization procedure. The resulting
enthalpy thus contains factors dependent on this acceleration parameter:
\be
M = \frac{m}{K\Xi} \sqrt{(\Xi+a^2/\ell^2)(1-A^2 \ell^2\Xi)}
\ee
that then propagate throughout the expressions for thermodynamic
volume and length. The accelerating black hole also obeys a Smarr
relation \cite{Smarr:1972kt}
\be\label{Smarr}
M=2(TS+\Omega J-PV)+\Phi Q\,,
\ee
that does not contain any trace of acceleration or tension.

There seems to be puzzle therefore: The first law has full cohomogeneity,
yet the Smarr relation has no tension. Further, while the expressions obtained in 
\cite{Anabalon:2018qfv} are perfectly adequate for the implicit study of
black hole thermodynamics, the `chemical' nature of the black hole is
less transparent, and typically has to be studied numerically, and
parametrically in terms of the horizon radius $r_+$ that, for example,
tracks entropy. 
To elucidate the \emph{chemical} nature of the black hole, and
to allow a more general analytic analysis of the phase space
our aim is therefore to have closed-form expressions, such as 
\eqref{ChrisRuf}, i.e.\ an integral expression of the form 
$M^2(S,P,Q,J,\mu_\pm)$,
together with expressions for the chemical
potentials in the form $\phi_i = {\partial M}/{\partial q_i}$,
where $q_i$ stands for a charge, $S, P, Q, J, \mu_\pm$ and 
$\phi_i$ its corresponding potential $T, V, \Phi, \Omega, \lambda_\pm$.
Given that the Smarr relation is a statement about scaling dimension,
and is given in terms of charges and potentials, it does not preclude
an expression for $M$ that includes the tensions.

The new physics in the accelerating black hole is that
of the conical deficit(s), and while the individual tensions
are natural geometric variables, they do not distinguish between 
an overall conical deficit, such as the cosmic string threading a black
hole (that does not have issues with a slow acceleration limit) and a 
differential conical deficit that produces a nett force on the
black hole, inducing acceleration. From the perspective of black hole
chemistry, it turns out that the conical deficits are more conveniently 
encoded in the average and differential conical deficits
of the spacetime:
\be
\beal
\Delta &= 1-2(\mu_+ + \mu_-) = \frac{\Xi}{K} \\
C &= \frac{(\mu_- - \mu_+)}{\Delta} = \frac{mA}{K\Delta} = \frac{mA}{\Xi}
\eeal
\label{deltadefs}
\ee
The tensions are bounded below by requiring positivity of energy
(or tension) and above by the fact that the maximal conical deficit is $2\pi$.
With $A\geq0$, so that $0\leq\mu_+\leq \mu_-\leq1/4$, this
translates into $0\leq C \leq \text{Min} \left \{ \frac{1}{2} ,
\frac{1-\Delta}{2\Delta}\right\}$. Although $\Delta$ and $C$ are
not unconstrained -- introducing an acceleration
necessarily also introduces an overall average deficit -- it proves to
be the best way to express the impact of the conical deficit on
the thermodynamics. Often, when considering an accelerating black
hole, the deficit on one axis (here $\mu_+$) is set to zero, in this case
$C=(1-\Delta)/2\Delta$, thus the upper bound is saturated and $\Delta
\in[\frac12,1]$. We are interested more generally in how conical deficits
impact thermodynamics, so will keep $C$ and $\Delta$ arbitrary, within 
their allowed ranges.

Now turn to the mass formula \eqref{ChrisRuf}. A check of the 
thermodynamic expressions in \cite{Anabalon:2018qfv} shows that 
$M, S$ and $Q$ all scale as $1/K$, and $J$ as $1/K^2$. This suggests 
that scaling each by $1/\Delta$ (or $\Delta^{-2}$ in the case of J)
is a promising starting point.
Some manipulations then reveal the appropriate remaining modifications,
and give the mass formula
\be
\beal
M^2 =
\frac{\Delta S}{4\pi}
\Bigl[ \left(1+\frac{\pi Q^2}{\Delta S}+\frac{8PS}{3\Delta}\right)^2
+\left(1+\frac{8PS}{3\Delta}\right)
\left(\frac{4\pi^2 J^2}{(\Delta S)^2} - \frac{3C^2 \Delta}{2PS}\right)
\Bigr]
\eeal 
\label{ChRu}
\ee
It is then a matter of algebra to confirm that the thermodynamic 
potentials conjugate to the charges, $T = \frac{\partial M}{\partial S} 
\big |_{P,J,Q,\mu_\pm}$ etc.\ correspond to the expressions in 
\cite{Anabalon:2018qfv} 
and are:
\be
\beal
V &=
\frac{2S^2}{3\pi M} \left[ \left(1+\frac{\pi Q^2}{\Delta S}+\frac{8PS}{3\Delta}\right)
+ \frac{2\pi^2J^2}{(\Delta S)^2} +\frac{9C^2\Delta^2}{32P^2S^2}
\right] \,,\\
T &= \frac{\Delta}{8\pi M}
\Bigg[\left(1+\frac{\pi Q^2}{\Delta S}+\frac{8PS}{3\Delta}\right)
\left(1-\frac{\pi Q^2}{\Delta S}+\frac{8PS}{\Delta}\right )
-\frac{4\pi^2 J^2}{(\Delta S)^2} -4C^2
\Bigg]
\,,\\
\Omega &= \frac{\pi J}{S M \Delta} \left(1+\frac{8PS}{3\Delta}\right) \,,\\
\Phi &= \frac{Q}{2M}\left(1+\frac{\pi Q^2}{S\Delta}+\frac{8PS}{3\Delta}\right)\,,\\
\lambda_\pm    &=
\frac{S}{4\pi M}
\Bigg[ \! \!\left ( \!\frac{8PS}{3\Delta} + \frac{\pi Q^2}{\Delta S} \!\right)^2
\!\!+ \frac{4\pi^2 J^2}{(\Delta S)^2} \left (\!1+ \frac{16PS}{3\Delta}\!\right)
\!-\left ( 1 \mp 2C \right)^2 \pm \frac{3C\Delta}{2PS}\!
\Bigg]
\,.
\eeal
\label{TDvars}
\ee
Since everything is now written in terms of the charges, this
elucidates the `chemical' structure of the accelerating black hole, and allows
for a more intuitive and natural analysis of the thermodynamics, as well
as clarifying some of the new phenomenology of accelerating thermodynamics.
We will now illustrate this by making some general observations on
the impact of conical deficits, before concluding by presenting a new
entropy bound for black holes with conical deficits.

The conical structure of the spacetime appears in two ways:
the `overall' conical deficit, encoded in $\Delta$, that can be present 
whether or not there is acceleration. $\Delta<1$ means that the 
spacetime contains a conical deficit, and if $C=0$, the deficit 
cuts through the whole spacetime, piercing the black hole. 
Acceleration appears via the parameter $C$. This is now more
interesting, as, unlike angular momentum and charge, that contribute
to the enthalpy positively, $C$ contributes negatively, indicating
an exothermic nature to this particular property. This now opens the possibility
of new phenomena in phase space, as, apart from the extremal limit, 
$T\to0$, we also potentially have a limit $M\to0$ that signals a breakdown 
in the thermodynamical description. This breakdown occurs approximately, 
though not precisely, at the breakdown of the slow acceleration r\'egime.

Let us begin by exploring the impact of an overall conical deficit,
setting $C=0$ and allowing $\Delta$ to vary. It might seem 
that as $\Delta$ simply enters as a rescaling parameter, it does not
change the qualitative thermodynamics, but the story is more subtle.
For example, consider the free energy $G=M-TS$, which in full is
\be
\beal
G = \frac{\Delta S}{8\pi M} \Biggl [ 
\frac{4 \pi^2 J^2}{\Delta^2 S^2} \left (3+\frac{16PS}{3\Delta} \right)
- 4 C^2 \left ( 1 + \frac{3\Delta}{4PS} \right)
+ \left (1 +\frac{8PS}{3\Delta} + \frac{\pi Q^2}{\Delta S}\right)
\left(1 -\frac{8PS}{3\Delta} + \frac{3\pi Q^2}{\Delta S}\right) \Biggr]\,.
\eeal
\label{gibbs}
\ee
For an uncharged, non-accelerating black hole, the magnitude of the 
free energy is decreased by adding a conical deficit. The 
Hawking-Page transition \cite{Hawking:1982dh} therefore still occurs
at $T_{HP} = \sqrt{8P/3\pi} = 1/\pi\ell$, and the critical point at which
the specific heat of the black hole becomes positive (i.e., the minimal 
temperature that a black hole can have) also remains at
$T_m = \sqrt{3} T_{HP}/2$, however, the free energy curves are strongly
modified with $\Delta$, and the entropy, or size, of the black hole
at each of these critical points is lowered: $S_c = \Delta/8P$, 
$S_{HP} = 3\Delta/8P$. With the addition of charge, the situation is
a bit more complicated, and depends on the interplay between the
charge and the deficit, but now, interestingly, the deficit causes the free
energy to increase in general (except for very small charge).

Now consider adding in acceleration, via the $C$ term. For large black
holes with or without charge, the presence of acceleration in itself
does not impact strongly on the thermodynamics, rather, it is the fact that
acceleration requires an average deficit that modifies the 
thermodynamics. However, as the charge, or size of the black hole
drops, things become much more interesting.
From \eqref{gibbs}, this clearly lowers the free energy, and for sufficiently
low charge and overall deficit, can eliminate any regions of positive $G$.
The lack of a Hawking-Page transition (for the simple reason that there 
is no spacetime with `half' a cosmic string without a black hole) was 
discussed in \cite{Appels:2016uha}. 
Although this global phase transition does not exist, 
for charged black holes there exists a local second order phase transition, 
first noted for non-accelerating black holes in \cite{Chamblin:1999tk}. 
Generically, for fixed $J$ there are two values of $S$ at which $C_P$,
the specific heat at constant pressure, diverges, but at some $J$ 
these two points coalesce, giving criticality. In the 
non-accelerating case, by expanding analogous quantities to \eqref{TDvars} 
this critical point was seen to possess mean-field exponents 
\cite{Dolan:2012jh}, the behaviour of a Van-der-Waals gas 
\cite{Kubiznak:2012wp}. Given equations \eqref{TDvars}, 
it is straightforward to perform series expansions in much 
the same manner for the accelerating case, the result yielding the
identical Ising-like exponents.

With acceleration there is also now a new phenomenon, 
noted in \cite{Abbasvandi:2018vsh}, that occurs
precisely because of the exothermic nature of acceleration. Recall
that the third law usually provides a lower bound on the entropy, 
corresponding to the size of the extremal black hole at which $T=0$.
However, in the presence of acceleration, we have a new limit coming from 
the positivity of $M^2$. To explore this, abbreviate notation by writing
\be
x = \frac{8PS}{\Delta} \;,\;\;\;\;
\frac{\pi Q^2}{\Delta S} = q \frac{C^2}{x} \;,\;\;\;
\frac{\pi J}{\Delta S} = j \frac{C^2}{x}
\ee
so that
\be
\beal
M^2 &= \frac{\Delta S}{4\pi} \biggl [ 
\left(1 + x -  q_+ \frac{C^2}{x} \right) \left (1 +x - q_-\frac{C^2}{x}\right)
+ 4 j^2 C^4 \frac{(1+x)}{x^2} \biggr]
\eeal
\ee
where $q_\pm = 2 - q \pm 2 \sqrt{1-q}$.
If $q\leq1$, then the roots $q_\pm$ are real, and there is a range of $j$ for which 
$M \to 0$ at some $x_0$. Further, since $T = \frac{1}{2M} 
\frac{\partial M^2}{\partial S}\propto  \frac{1}{M} \frac{\partial M^2}{\partial x}$,
this occurs before the extremal limit is reached. For this range
of low charge/rotation to acceleration ratios, small black holes are no longer
cold, but instead, like their uncharged counterparts, are hot, and
have a negative specific heat 
as the enthalpy tends to zero. As the charge/rotation increases, a critical
limit is reached, $q_c(j)$ for which $M^2$ has a repeated zero, at which
$T$ is finite, and above the critical values of charge/rotation, the black hole
once again exhibits a swallowtail. This `snapping' of the swallowtail was 
discovered in \cite{Abbasvandi:2018vsh}, although the snapping point 
could not be determined analytically from the implicit expressions
of the thermodynamical variables. 

Using the chemical variables, we can readily find the one-parameter family of
critical charged, rotating, and accelerating black holes that have
infinite enthalpy, but finite temperature at the snapping point of
the swallowtail. These critical entropies and temperatures 
are given implicitly by the value of $x$
at which the mass becomes zero, $x_0$, and $C$:
\be
\beal
Q_{_\text{S}}^2 &= \frac{3\Delta^2}{8\pi P}(1+x_0) 
\left[1+x_0 - \sqrt{(1+2x_0)^2-4C^2}\right]\\
J_{_\text{S}}^2 &= \frac12 \left (\frac{3\Delta^2}{8\pi P} \right)^2 (1+2x_0)
\Bigl [ 2C^2 
- (1+x_0) \left ( 1+2x_0 - \sqrt{(1+2x_0)^2-4C^2} 
\right)\Bigr] \\
T_{_\text{S}} &= \sqrt{\frac{2P}{3\pi x_0}} \Bigl[ (4x_0+3)(2x_0+1)-4C^2 
-2(1+x_0) \sqrt{(1+2x_0)^2-4C^2} \Bigr]^{1/2}
\eeal
\ee
where $x_0\in [\frac{\sqrt{1+4C^2}-1}{2}, \frac{\sqrt{1+12C^2}-1}{3}]$. 
The lower limit has $J=0$, $Q^2 =\delta\mu^2\ell^2$, 
as noted in \cite{Abbasvandi:2018vsh}, and the upper limit corresponds to 
$Q=0$, $J= \Delta^2 \ell^2 x_0\sqrt{2x_0+1} /2$, where $x_0$ takes
the upper limit value. What is interesting here is how the overall deficit
plays a role in the critical values of $Q$ and $J$, except for the pure charge
snapped swallowtail, where it seems to be the acceleration that is primary
driver. As the acceleration decreases, $C\to0$,  and the range and size of $x_0$
correspondingly decreases, thus the critical temperature for the snapping
point also becomes higher, with the critical temperature increasing as $x_0$
moves towards the lower end of the range (zero angular momentum).
The maximum value of acceleration, $C=1/2$, corresponds to a deficit of 
$2\pi$ along the South axis, and has the lowest values of critical
temperature and the largest range of $q$ and $j$. The absolute lowest critical
snapping temperature occurs for $Q=0$, and is $T = \sqrt{2}/\pi$.

To conclude our exploration of the accelerating black hole
chemistry, consider
the \emph{Reverse Isoperimetric Inequality} \cite{Cvetic:2010jb}.
The Isoperimetric Inequality is the simple geometric statement
that the largest surface area enclosed by a loop of string is when 
the string is circular, or its higher dimensional equivalent:
$ ({\cal A}/{\cal A}_0)^D \geq ( {\cal V}/{\cal V}_0)^{D-1}$,
where the subscript $0$ indicates the volume or area of a unit sphere.
However, for the black hole, the area directly determines the entropy,
thus we would expect that for a given volume, a black hole would want
to maximise its entropy, which runs counter to the standard inequality that
would mean a spherical black hole would be the lowest entropy for that
volume, and hence unstable.
Analysing a wide range of solutions, Cvetic et al.\ \cite{Cvetic:2010jb}
discovered that black holes always satisfied the \emph{inverse} of
this inequality:
\be
\left (\frac{{\cal A}}{{\cal A}_0}\right)^D \leq 
\left ( \frac{{\cal V}}{{\cal V}_0}\right)^{D-1}
\label{reviso}
\ee
leading to their \emph{Reverse Isoperimetric Inequality} conjecture.

Let us now explore this inequality in the context of black holes with
conical deficits. Note that
\be
\beal
\frac{4 \pi M^2}{\Delta S} &= \left ( 1+ \frac{\pi Q^2}{\Delta S}+ x \right)^2 
+ 4\left ( 1 + x \right) \left ( \frac{\pi^2J^2}{\Delta^2S^2}- \frac{C^2}{x}\right)\\
&= \left ( \frac{3\pi M V}{2 S^2} - \frac{2C^2}{x^2} \right)^2
- 4 \left (\frac{\pi Q^2}{\Delta S}\right) \left (\frac{\pi J}{\Delta S}\right)^2 
-4 \left (\frac{\pi J}{\Delta S}\right)^4 - 4(1+x)\frac{C^2}{x} 
\eeal
\ee
At this point, we spot that the term in brackets on the RHS
contains the seed of the isoperimetric ratio:
\be
M^2 \left ( \frac{3V}{4\pi} \right)^2 \left ( \frac{\pi}{S} \right)^4
\geq \left ( \frac{3\pi M V}{4 S^2} - \frac{C^2}{x^2} \right)^2
\geq \frac{\pi M^2}{\Delta S}
\ee
from which we may conclude a \emph{new Reverse Isoperimetric
Inequality}, appropriate for spacetimes with a conical deficit:
\be
\left ( \frac{3V}{4\pi} \right)^2 \geq \frac{1}{\Delta}  
\left ( \frac{{\cal A}}{4\pi} \right)^3
\ee
with equality iff $C = J =0$.
The larger the conical deficit, the smaller the entropy is with respect to the 
volume, thus conical defects appear to render black holes sub-entropic.

We have highlighted here some key distinct phenomena displayed 
by the deficited black hole, and in a companion paper we explore the 
properties of conical holographic heat engines \cite{inprep}.
Apart from the interesting new phenomenology displayed by accelerating
black holes, the ease with which we are able to analyse and uncover
this behaviour using the chemical variables is remarkable. Not only have
we proved a new Entropy Inequality for black holes with conical deficits,
but by writing down a Christodoulou-Ruffini type of mass formula, we
have now expressed the mass of a four-dimensional black hole that
includes two new charges: $\Delta$ and $C$, or, the average deficit and
acceleration. It would be interesting to understand how our expressions
relate to those in \cite{Astorino:2016hls} for an unconfined magnetic flux 
tube through a black hole. 
One possibly disturbing aspect of the exothermic impact of
acceleration is that it seems to diverge as the pressure tends to zero. This
is a feature of the slow-acceleration limit, or, the fact that the time coordinate
of a slowly accelerating black hole is proportional to the time coordinate of
an asymptotic observer, thus the notion of an isolated mass makes 
sense. For a black hole whose time coordinate is a boost coordinate,
such as the usual vacuum accelerating black hole or Rindler spacetime, 
alternate thermodynamic parameters, along the lines of the
boost mass developed in \cite{Dutta:2005iy} should be used. Clearly, the 
thermodynamics and chemistry of these new charges for a black
hole merits further study and exploration.

\SEC{Acknowledgments.}
It is a pleasure to thank David Kubiz\v n\'ak for many interesting discussions.
RG is supported in part by the STFC [consolidated grant ST/P000371/1], 
and in part by the Perimeter Institute. AS is supported by an STFC 
studentship, and would also like to thank Perimeter Institute for 
hospitality while this research was undertaken. 
Research at Perimeter Institute is supported by the Government of Canada 
through the Department of Innovation, Science and Economic Development 
and by the Province of Ontario through the Ministry of Research and Innovation.

\end{document}